\documentclass[aps,pre,twocolumn,showpacs,superscriptaddress,amsmath,amssymb,longbibliography]{revtex4-2}

\usepackage[normalem]{ulem}
\usepackage{graphicx}
\usepackage{color,soul}
\usepackage{float}
\usepackage{mathtools}
\usepackage{hyperref}
\usepackage{xcolor}

\makeatletter
\def\p@figure{\color{blue}}
\def\p@equation{\color{blue}}
\def\p@table{\color{blue}}
\def\p@bibliography{\color{blue}}
\makeatother

\graphicspath{ {./Figures/} }

\begin{document}
\title{3D model for surface accumulation of chiral and non-chiral microswimmers}

 \author{Danne M. van Roon}
  \affiliation{Centro de F\'isica Te\'orica e Computacional, Faculdade de Ci\^encias, Universidade de Lisboa, 1749-016 Lisboa, Portugal }
  \affiliation{Departamento de F\'isica, Faculdade de Ci\^encias, Universidade de Lisboa, 1749-016 Lisboa, Portugal}
  \author{Giorgio Volpe}
  \affiliation{Department of Chemistry, University College London, 20 Gordon Street, London WC1H 0AJ, UK. }
  \author{Margarida M. Telo da Gama}
  \affiliation{Centro de F\'isica Te\'orica e Computacional, Faculdade de Ci\^encias, Universidade de Lisboa, 1749-016 Lisboa, Portugal }
  \affiliation{Departamento de F\'isica, Faculdade de Ci\^encias, Universidade de Lisboa, 1749-016 Lisboa, Portugal}
  \author{Nuno A. M. Ara\'ujo}
  \affiliation{Centro de F\'isica Te\'orica e Computacional, Faculdade de Ci\^encias, Universidade de Lisboa, 1749-016 Lisboa, Portugal }
  \affiliation{Departamento de F\'isica, Faculdade de Ci\^encias, Universidade de Lisboa, 1749-016 Lisboa, Portugal}

\begin{abstract}
Persistent motion of microswimmers near boundaries is known to result in surface accumulation. Recently it was shown experimentally that surface accumulation of microswimmers is impacted primarily by steric forces and short-ranged hydrodynamics. A way to control surface accumulation is by reducing the contact surface area between swimmers and surface by modifying its topography, typically through the application of microscale structure. In this work, we introduce a three-dimensional(3D) model of a microswimmer navigating a volume bounded by a top and bottom surface. We describe the swimmer-surface interaction with an effective short-ranged hydrodynamic alignment force, and study numerically the effect of surface textures, modelled by convex obstacles, on the surface accumulation of chiral and non-chiral microswimmers. We find that, depending on the angular velocity of the swimmer, and the alignment force, convex obstacles can either hinder or enhance surface accumulation. We discuss potential applications to sorting of microswimmers by their angular velocity.
\end{abstract}

\maketitle
\section{Introduction}
In the last two decades, active matter has become an increasingly important focus of research \cite{roadmap, space}. On small length scales, active matter describes micron-sized entities such as algi, sperm cells, bacteria and artificial microswimmers including self propelled Janus particles and active droplets \cite{rev}. An important characteristic of microswimmers is their tendency to accumulate at surfaces \cite{rev, steering_confinement, hydro, accumul}. Steric effects and the persistence in the swimming direction are generic contributing factors to  surface accumulation (or ‘trapping’), in addition to hydrodynamic effects that depend on the specifics of the swimmer and the surface \cite{ following1, following2, following3}. For example, driven by hydrodynamic interactions, several microswimmers move in chiral trajectories, typically along circles, when swimming near a flat interface \cite{whychiral, whychiral2, whychiral3, rev_liebchen}.

Bacterial surface dynamics are central in various industrial, biomedical, and environmental processes  \cite{optimal, biofouling1, biofouling2}. On the one hand, the adhesion of bacteria to surfaces frequently results in the formation of persistent biofilms that are difficult to remove, causing challenges in various fields, including fouling of water purification systems \cite{copper}, corrosion of structures used to transport and store chemicals \cite{industrial}, and adhesion to medical implants, where bacterial infection can result in inflammation that can even lead to death \cite{medical}. On the other hand, the industrial potential of biofilms is becoming increasingly developed, including biorefineries \cite{biorefineries}, bioremediation to remove contaminants from freshwater and wastewater \cite{bioremediation}, and as templates for new materials with applications in construction and industry \cite{template}.

A quantitative understanding of surface entrapment and subsequent adhesion could further the development of engineered materials to control and prevent bacterial adhesion to surfaces \cite{surface1, surface2, surface3}. In recent years, the effects of surface topography and roughness on bacterial surface dynamics and adhesion have received increasing attention \cite{surface_rev, near_surfaces_rev}. Experimental observations show that the topography of the environment can strongly influence the dynamics of microswimmers on a surface, in non-intuitive ways. Experimental evidence indicates that the presence of porous micro-structures generally hinders the diffusive transport of microswimmers \cite{experiments1, lorentzgas}. Interestingly, for chiral microswimmers, contrasting phenomenology has also been observed. For example, a significantly enhanced propagation on surfaces, due to randomly placed obstacles, has been reported in theoretical studies \cite{franosh1, franosh2, chiral} and in experiments with \textit{E. coli} \cite{forward}. Furthermore, experiments tracking \textit{E. coli} navigating a colloidal crystal revealed that the colloids rectify the trajectories of the bacteria, resulting in enhanced transport \cite{lattice, lattice2, lattice3, 3d}.
\begin{figure*}[tp]
     \centering
     \includegraphics[width=17 cm]{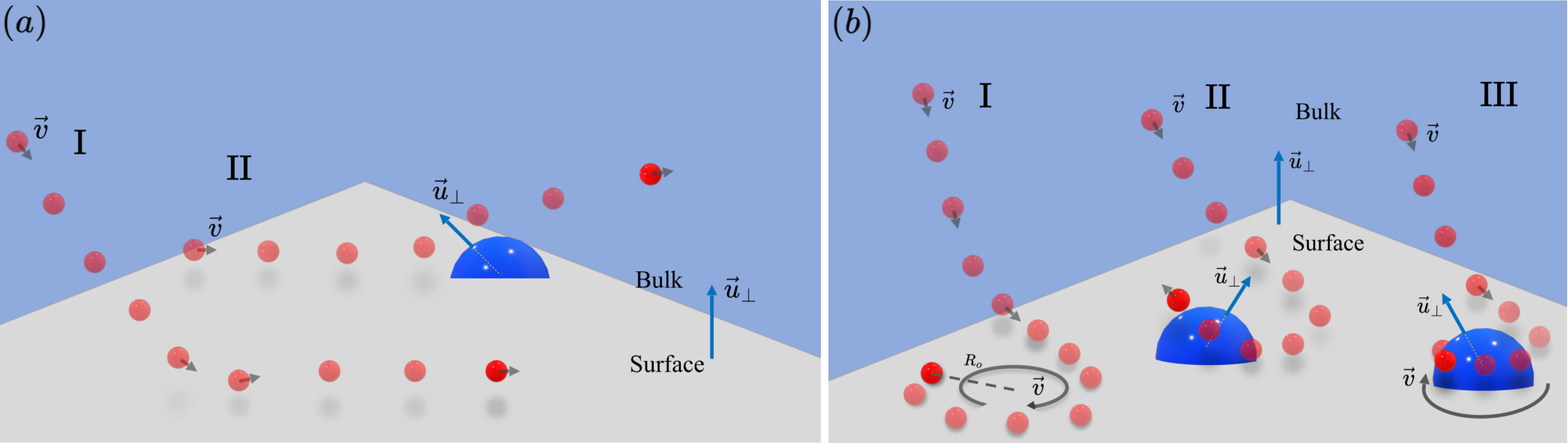}
       \caption{(a,b) Schematic depiction of a swimmer (red) moving at velocity $\vec{v}$ in the volume (blue) near the surface (light grey). The dark grey arrows indicate the direction of motion of the swimmer, the blue arrows indicate the vector perpendicular to the closest surface plane, or the boundary of the closest obstacle (dark blue hemisphere). (a) A non-chiral swimmer (I) approaching the surface, (II) guided away from the surface after interacting with the obstacle. (b) A chiral swimmer ($\Omega \neq 0$) (I) approaching the surface, where it moves in orbits of radius $R_{\textrm{o}}$, (II) guided away from the surface after interacting with the obstacle, (III) trapped in an orbit around the boundary of the obstacle.}
        \label{fig_schematic}
\end{figure*}

Recently experiments on \textit{E. coli} approaching a surface, have shown that the average reorientation of the cells in a direction parallel to the surface is driven by steric forces at contact and short-ranged hydrodynamics, with long-range hydrodynamics playing only a minor role \cite{leonardo1, leonardo2, shortranged, surface_rev}. Similar results were obtained for sperm cells on a surface \cite{bullsperm}, for \textit{E. coli} interacting with microsized pillars \cite{following3} and for synthetic microswimmers navigating an environment of passive colloidal beads \cite{following1}. After the reorientation event, a swimmer was observed to move along a boundary (surface, pillar or bead) until Brownian diffusion rotates its axis away.

In this article we introduce a 3D model to examine the surface accumulation of microswimmers induced by steric forces and an effective short-ranged hydrodynamic alignment force that aligns the propagation direction of the swimmers along the nearest boundary. We take \textit{E. coli} and bull sperm cells as examples, navigating a volume bounded by a top surface plane and a bottom one, akin to a microfluidic channel. We study the effect of placing convex obstacles, which are three times the size of the swimmers, randomly on both surfaces. We examine how the surface accumulation is impacted by the strength of the alignment force and the angular velocity of the motion. We quantify the surface accumulation by measuring the fraction of swimmers near the surfaces. We find that the presence of obstacles always reduces the accumulation of non-chiral microswimmers on these surfaces, as, by aligning along the boundary of the obstacles, the swimmers are directed away from the surface.
As the obstacle density is increased, the surface accumulation is reduced. For chiral microswimmers, the accumulation can be reduced or enhanced. As previously observed in \cite{following1,following2,following3}, we confirm with our model that a chiral swimmer may get trapped. This orbital trapping results from the chiral motion of the swimmer, and only occurs if the obstacles are at least equal to the size of the swimming orbit. We discuss how obstacles can be used to control the accumulation of chiral swimmers near surfaces, by carefully tuning their size. In addition, we discuss how obstacles can be employed to sort swimmers based on their angular velocity.  

\section{Model}
We consider a spherically shaped swimmer of diameter $\sigma$ (for natural and synthetic microswimmers typically 1-5 \textmu m) moving with velocity $\vec{v}$ and corresponding momentum $\vec{p}$. The swimmer navigates a volume of thickness 100$\sigma$ bounded by a top and bottom square surface plane, with edge size $L$ and periodic boundary conditions along the directions parallel to the surfaces. Each surface plane is covered with $N_{\textrm{o}}$ non-overlapping obstacles of diameter $\sigma_{\textrm{o}} = 6\sigma$, distributed uniformly at random. The obstacles are quantified by the surface coverage defined as: $\rho = \frac {\mathrm{N_{\textrm{o}}}\pi{\sigma_{\textrm{o}}}^2}{4L^2}$ ($\times 100 \%$). The motion of the swimmer is determined by the force acting on it, and a stochastic term $\vec{\xi}$ that is parameterized to set a proper rotational diffusion constant for the swimmer. The force has two contributions: $\vec{F}_{\textrm{motile}}$ responsible for the self-propelled motion in the volume and $\vec{F}_{\textrm{surface}}$ describing the interaction with the surface. The trajectory of the swimmer is obtained by integrating the following equation
\begin{equation}
\dot{\vec{p}} = \vec{F}_{\textrm{motile}} + \vec{F}_{\textrm{surface}} + \vec{\xi}.
\end{equation}
The motile force is
\begin{equation}
\vec{F}_{\textrm{motile}} = \frac{1}{\tau}(\vec{p}-p_0\hat{p}),\  
\end{equation}
which drives the motion along a predefined direction $\hat{p}$, with a momentum of magnitude $p_0$. The timescale of the motility is set by $\tau$. 
For the surface interaction we take
\begin{equation}\label{surface}
\vec{F}_{\textrm{surface}} =-\alpha(\hat{p}\cdot\hat{\mu}_{\perp} )\hat{\mu}_{\perp} + \beta(\hat{p}\times\hat{\mu}_{\perp} ) +  \vec{F}_{\textrm{steric}},
\end{equation}
where the first term is an effective alignment force between the swimmer and an obstacle or surface plane, the second term is the angular velocity of the motion when the swimmer approaches the surface, and the last term is the steric interaction of the swimmer with the obstacles and the surface. The strength of the alignment interaction is determined by $\alpha$. For $\alpha > 0$ the force will align the swimmer along the obstacle boundary or surface plane, with $\hat{\mu}_{\perp}$ the unit vector perpendicular to the obstacle boundary or surface plane closest to the swimmer \cite{leonardo1, leonardo2}.

For $\beta > 0$, the swimmer will trace circular trajectories when moving close to the surface. The orbital radius of the trajectories $R_{\textrm{orbit}}$ is determined by $\beta$ and defined as $R_{\textrm{orbit}} = \frac{v}{\beta}$\cite{whychiral, whychiral2, whychiral3, rev_liebchen}. We introduce $\Omega = \frac{\sigma_{\textrm{o}}}{2 R_{\textrm{orbit}}}$ as a measure of the angular velocity, by normalizing the size of the obstacles with the size of the swimming orbit. For larger values of $\beta$ swimmers trace smaller orbits corresponding to larger values of $\Omega$. Experimentally, $\Omega$ could be controlled through the obstacle size, with larger obstacles resulting in larger $\Omega$ \cite{obstacle_size, following3}.
A schematic illustration of a non-chiral swimmer ($\beta = 0$) is shown in Fig. \ref{fig_schematic} (a,I), and an illustration of a chiral swimmer in Fig. \ref{fig_schematic} (b,I). To represent the short-ranged nature of the interactions, a cut-off distance $r_{\textrm{c}} =\frac{3\sigma}{2}$ is introduced, such that, when the distance $r$ between swimmer and the obstacle boundary or surface plane is $r > r_{\textrm{c}}$, $\alpha$ and $\beta$ are set to 0.

The last term in eq. \ref{surface} is a steric interaction of the swimmer with the obstacles and the surface. The interaction with the surface is modelled by an exponential repulsion of magnitude
\begin{equation}\label{steric}
F_{\textrm{steric}}=\frac{1}{r_{\textrm{s}}}\exp{(-r_{\textrm{s}})},
\end{equation}
with $r_{\textrm{s}}$ the distance between swimmer and surface. The interaction becomes effective when $r_{\textrm{s}} < \frac{\sigma}{2}$. For the interaction between the swimmer and the obstacles a truncated Weeks-Chandler-Anderson potential is used \cite{chiral}. The time evolution is obtained by integrating the system with the velocity Verlet method. In the following we will express distances in terms of the dimensionless swimmer radius $\sigma/2$ and time in terms of $\sigma/v$. A simulation typically includes $N_{\textrm{o}} = 100$ obstacles on each surface plane, for a total of 200 obstacles. Finally, the step size in the simulation is $\Delta t = 10^{-4}$, $\tau=1$ and a simulation lasts for $t$ = 3600 $\sigma/v$.

\begin{figure*}[tp]
     \centering
     \includegraphics[width=17.1 cm]{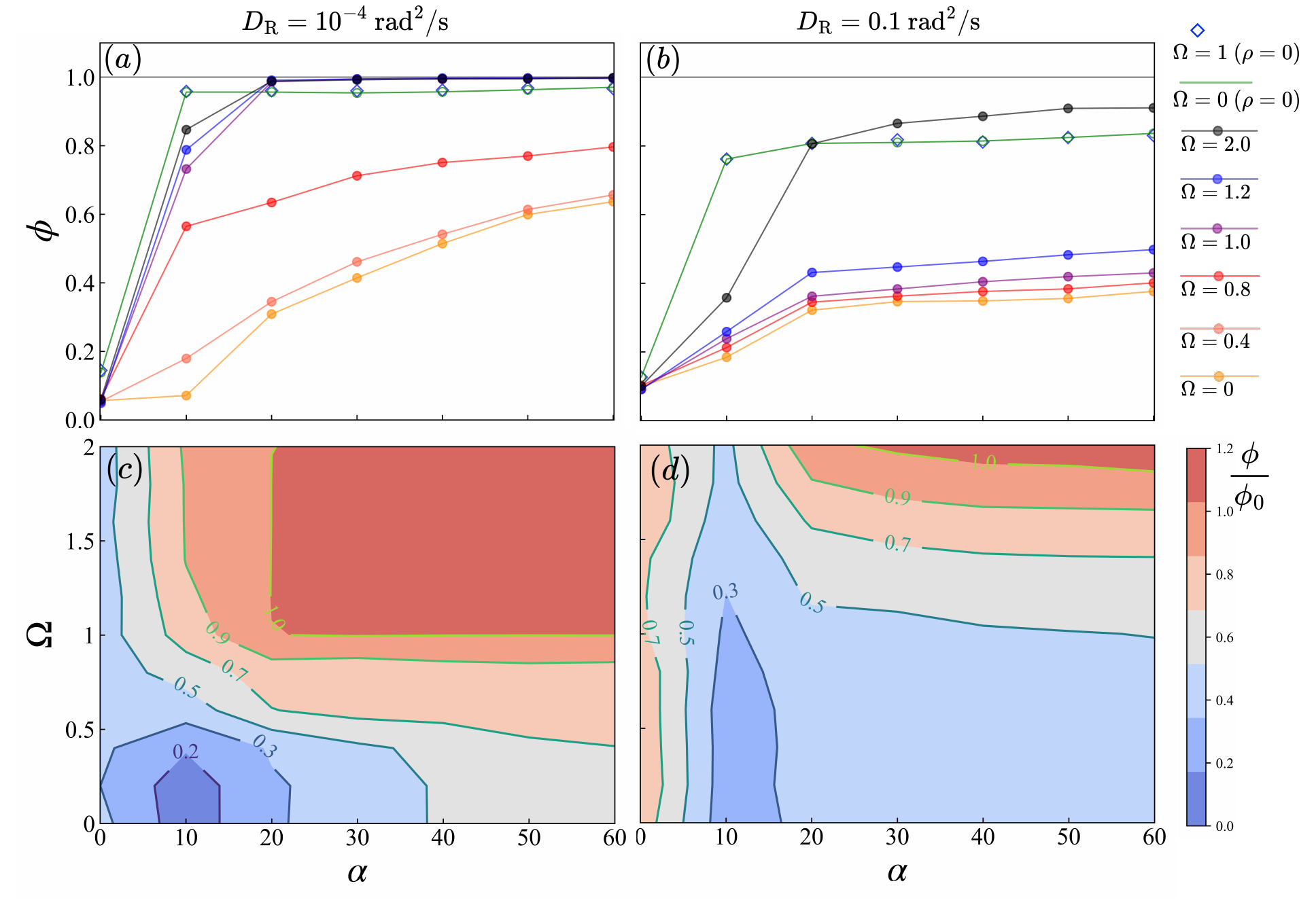}
        \caption{(a,b) Fraction of swimmers accumulated on the surface $\phi$ against the alignment force $\alpha$ for different angular velocities $\Omega$, with an obstacle surface coverage $\rho = 20 \%$: (a) for swimmers with $D_{\textrm{R}}=10^{-4}$ $\mathrm{rad}^2/\mathrm{s}$ and (b) for swimmers with $D_{\textrm{R}}=0.1$ $\mathrm{rad}^2/\mathrm{s}$. (c,d) Fraction $\frac{\phi}{\phi_0}$ of swimmers accumulated on a surface with convex obstacles relative to the fraction accumulated at a smooth surface $\phi_0$, as a function of $\Omega$ and $\alpha$, (c) for $D_{\textrm{R}}=10^{-4}$ $\mathrm{rad}^2/\mathrm{s}$ and (d) for $D_{\textrm{R}}=0.1$ $\mathrm{rad}^2/\mathrm{s}$.}
        \label{fig_phi}
\end{figure*}

\section{Results}
To characterize the surface accumulation, we count the number of swimmers $N_{\textrm{surface}}$ that are at a distance shorter than $3\sigma$ (one obstacle radius) from the bottom or top surfaces in the steady state of the system. We define the fraction $\phi$ of microswimmers accumulated at the surfaces as,
\begin{equation}\label{fraction}
\phi = \langle \frac{N_{\textrm{surface}}}{N} \rangle.
\end{equation}
The brackets $\langle . \rangle$ indicate an ensemble average over $N=25000$ swimmer trajectories (the swimmers do not interact), each beginning at a position selected uniformly at random within the volume. For smooth surfaces $\rho=0$, the surface planes contain no obstacles. For surfaces with obstacles, each surface plane is covered with $N_{\textrm{o}}$ non-overlapping hemispherical obstacles, uniformly and randomly placed on the surface, such that their centres are on the surface plane. Results of 250 swimmer trajectories are averaged over 100 obstacle configurations, for a total $N=25000$ swimmer trajectories. Two different values for the rotational diffusion will be considered, $D_{\textrm{R}}=0.1$ $\mathrm{rad}^2/\mathrm{s}$ corresponding to the rotational diffusion measured for \textit{E. coli} \cite{shortranged}, and $D_{\textrm{R}}=10^{-4}$ $\mathrm{rad}^2/\mathrm{s}$ corresponding to the value for bull sperm \cite{bullsperm}.

\subsection{Smooth surface}
To set the stage first consider the simplest case of non-chiral ($\Omega = 0$) swimmers navigating a volume bounded by smooth surfaces, where the surface accumulation $\phi$ increases with the alignment force strength $\alpha$. Fig. \ref{fig_phi} (a,b) (blue diamonds) shows how the surface accumulation $\phi$ increases with $\alpha$ ranging from $\alpha=0$, where the rotational diffusion governs the dynamics, to $\alpha=60$, where the alignment force dominates. 

In the absence of alignment force ($\alpha = 0$), about 15$\%$ ($D_{\textrm{R}}=10^{-4}$ $\mathrm{rad}^2/\mathrm{s}$, Fig. \ref{fig_phi} a ) and 12$\%$ ($D_{\textrm{R}}=0.1$ $\mathrm{rad}^2/\mathrm{s}$ Fig. \ref{fig_phi} b), swimmers accumulate near the surface. This accumulation is a result of the persistent motion of the swimmers. A swimmer explores the volume until it collides with one of the surfaces, where it will stay until rotational diffusion directs it away from it, and it can escape.  

As the alignment force increases, the fraction of swimmers accumulated near the surfaces increases. This behaviour can be explained by considering that the alignment force competes with the rotational diffusion, as a stronger tendency to align prevents the swimmer from orienting away from the surface and escape. The marked increase in accumulation for $0 < \alpha < 10$ results from this competition. For $D_{\textrm{R}}=10^{-4}$ $\mathrm{rad}^2/\mathrm{s}$ (Fig. \ref{fig_phi} a) when $\alpha \geq 10$ the alignment force becomes strong enough to trap the swimmer at the surface. The trapping results in an enhanced accumulation, which increases with the alignment force, reaching 95$\%$ for $\alpha \geq 10$. When $D_{\textrm{R}}=0.1$ $\mathrm{rad}^2/\mathrm{s}$ (Fig. \ref{fig_phi} b), the tendency of the swimmers to diffuse away from the surface is stronger, reducing the accumulation. The accumulation reaches 76$\%$ for $\alpha = 10$ continuing to increase slowly to 82$\%$ for $\alpha = 60$.

When chiral swimmers are considered ($\beta \neq 0$ and by extension $\Omega \neq 0$), the accumulation behavior is unchanged for smooth surfaces. In Fig. \ref{fig_phi} (a,b), the green line shows the accumulation for $\Omega=1$, which coincides with the blue diamonds for non-chiral swimmers ($\Omega=0$). The chirality causes the swimmers to trace circular trajectories along the surface, but does not affect their overall surface accumulation.

\subsection{Surface structured with convex obstacles}
A very different behavior is observed, for both values of $D_{\textrm{R}}$, when randomly placed obstacles are added to the surface. We begin by considering swimmers with $D_{\textrm{R}}=10^{-4}$ $\mathrm{rad}^2/\mathrm{s}$.

\begin{figure*}[tp]
     \includegraphics[width=17.5cm]{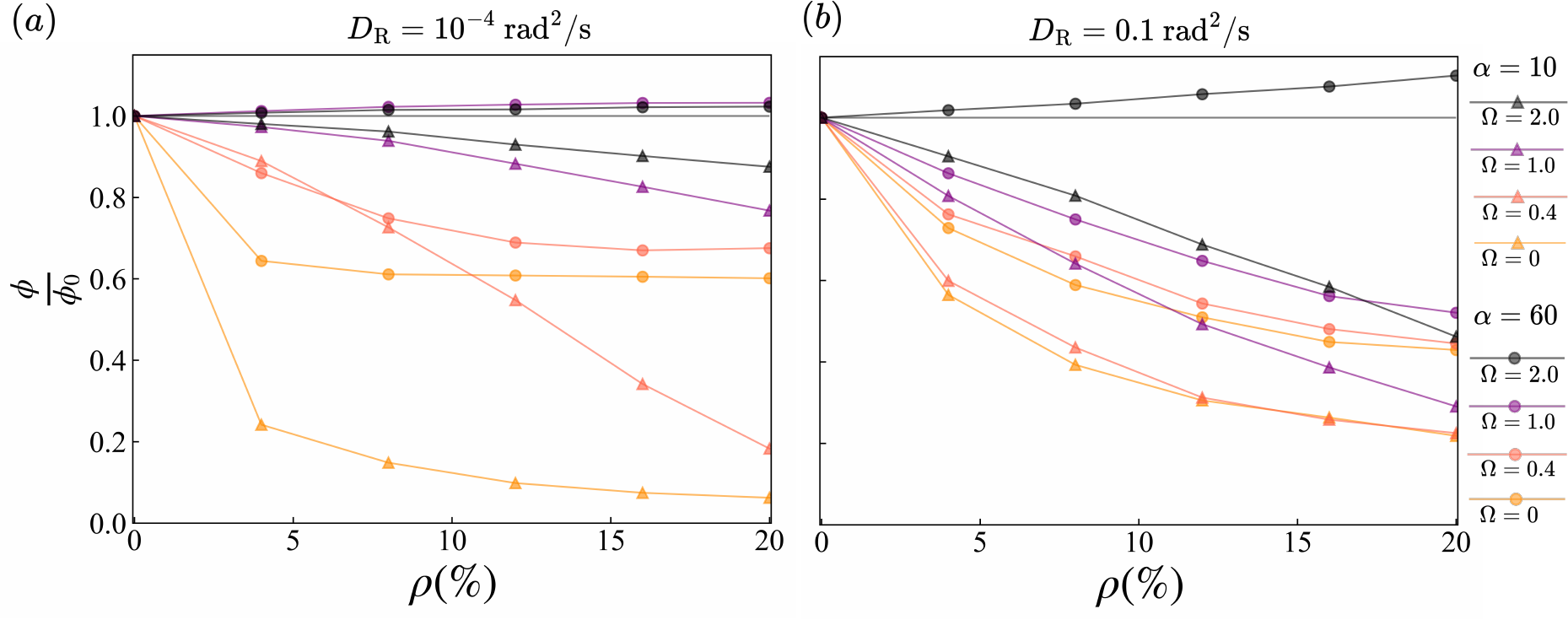}  
     \centering
        \caption{Fraction $\phi$ of swimmers accumulated near the surface for a given angular velocity $\Omega$ normalized to $\phi_0$, the accumulation near a smooth surface for the same angular velocity, against the obstacles density $\rho$. The rotational diffusion is (a) $D_{\textrm{R}}=10^{-4}$ $\mathrm{rad}^2/\mathrm{s}$ and (b) $D_{\textrm{R}}=0.1$ $\mathrm{rad}^2/\mathrm{s}$ with alignment force strengths $\alpha=10$ (triangles) and $\alpha=60$ (circles). } 
        \label{fig_rho}
\end{figure*}

For non-chiral and chiral swimmers in a volume bounded by surfaces with obstacle surface density $\rho = 20 \%$, the accumulation near the surfaces (lines with circles in Fig. \ref{fig_phi} a) is reduced for all $\alpha$ for $\Omega < 1$ and for $\alpha \leq 10$ for $\Omega \geq 1$. Fig. \ref{fig_phi} (c) shows the fraction $\frac{\phi}{\phi_0}$ of accumulated swimmers with angular velocity $\Omega$ near a surface with obstacles $\phi$ with respect to that of swimmers with the same angular velocity near a smooth surface $\phi_0$, as a function of the alignment force $\alpha$ and angular velocity $\Omega$.  

For non chiral swimmers and chiral swimmers with $0 \leq \Omega \leq 1$, the reduction of the surface accumulation by the obstacles ($\frac{ \phi }{\phi_0 }<1$ in Fig. \ref{fig_phi} c) is a result of the modified surface structure. In the absence of an alignment force ($\alpha = 0$), about 8$\%$ of the swimmers accumulate at the surface. This accumulation is a result of the persistent motion of the swimmers, and the decrease in accumulation relative to the smooth surface can be attributed to the reduction of the flat surface area available to swimmers, because of the presence of obstacles. For $\alpha > 0$, when approaching an obstacle, a swimmer aligns its direction of motion along the convex boundary of the obstacle, instead of the surface plane, so that its orientation is directed away from the surface. As $\alpha$ is increased, the tendency to align along the surface or boundary of an obstacle becomes stronger, increasing the surface accumulation. A schematic trajectory that leaves the surface after aligning along an obstacle is shown for a non-chiral swimmer in Fig. \ref{fig_schematic} (a,II) and a chiral swimmer in Fig. \ref{fig_schematic} (b,II). For $0 \leq \Omega < 1$ the accumulation increases with the angular velocity (Fig. \ref{fig_phi} a). This can be explained by the chiral swimmers exploring the surface less extensively, thereby encountering fewer obstacles when compared to less chiral swimmers. As the angular velocity increases, the swimming orbits become smaller, decreasing the efficiency with which the surface is explored. This limits the ability of the obstacles to reduce the accumulation at the surface and leads to an increase of $\frac{\phi}{\phi_0}$ with the angular velocity in Fig. \ref{fig_phi} (c).

For chiral swimmers with a larger angular velocity ($\Omega \geq 1$) a different behavior is observed, where swimmers move in persistent orbits along the boundary of an obstacle, until the rotational diffusion directs them away from the obstacle and they escape. A schematic example trajectory of a swimmer that orbits an obstacle is displayed in Fig. \ref{fig_schematic} (b, III). For $D_{\textrm{R}}=10^{-4}$ $\mathrm{rad}^2/\mathrm{s}$, the rotational diffusion is relatively weak and the swimmers can become effectively trapped at the obstacles for a sufficiently strong alignment force $\alpha \geq 20$, increasing the surface accumulation in Fig. \ref{fig_phi} (a). We note that due to this ``orbital trapping'', the ability of the obstacles to direct swimmers away from the surface is mitigated for $\alpha \geq 20$, whereby obstacles will rather enhancing surface accumulation ($\frac{\phi}{\phi_0} > 1$ in Fig. \ref{fig_phi} c). 

The trapping of a swimmer in an orbit around an obstacle is a result of the interplay between the angular velocity and the alignment force along the boundary of the obstacle. When a swimmer approaches an obstacle, the alignment force will guide the swimmer around its boundary. For a swimming orbit that is equal to the size of the obstacle ($\Omega= 1.0$), this results in a circular trajectory centered in the obstacle. When the swimming orbit is of the size, or smaller than the size of the obstacle, $(\Omega \geq 1)$, the angular velocity will continuously direct the swimmer towards the centre of the obstacle with the alignment force directing the swimmer back along the boundary. A larger angular velocity will result in a stronger tendency to align along the boundary. Detachment of a swimmer from an orbit can occur when the Brownian diffusion is effective in reorienting the swimmer to escape from its orbit.

For $D_{\textrm{R}}=0.1$ $\mathrm{rad}^2/\mathrm{s}$, the larger diffusion randomizes the motion, increasing the boundary detachment and reducing surface accumulation (Fig. \ref{fig_phi} b). By perturbing the circular trajectories the diffusion increases the surface exploration of chiral swimmers, reducing the difference between different angular velocities $\Omega$. Due to enhanced surface exploration, obstacles become more effective in reducing near surface accumulation of chiral swimmers, decreasing $\frac{\phi}{\phi_0}$ in Fig. \ref{fig_phi} (d) (when compared to $D_{\textrm{R}}=10^{-4}$ $\mathrm{rad}^2/\mathrm{s}$, in Fig. \ref{fig_phi} c) for all but $\alpha <10$ when the alignment force is weak and diffusion dominates. 

The effect of orbital trapping on surface accumulation is reduced significantly for most values of $\Omega$. The diffusion breaks the orbits reducing the surface accumulation promoted by the obstacles ($\frac{\phi}{\phi_0}<1$ in Fig. \ref{fig_phi} d) with the exception of $\Omega=2$, where the angular velocity is strong enough to keep the swimmer aligned along the boundary of an obstacle, effectively trapping it for $\alpha > 20$ ($\frac{\phi}{\phi_0}>1$ in Fig.\ref{fig_phi} d). 

\begin{figure*}[tp]
     \centering
     \includegraphics[width=17.5 cm]{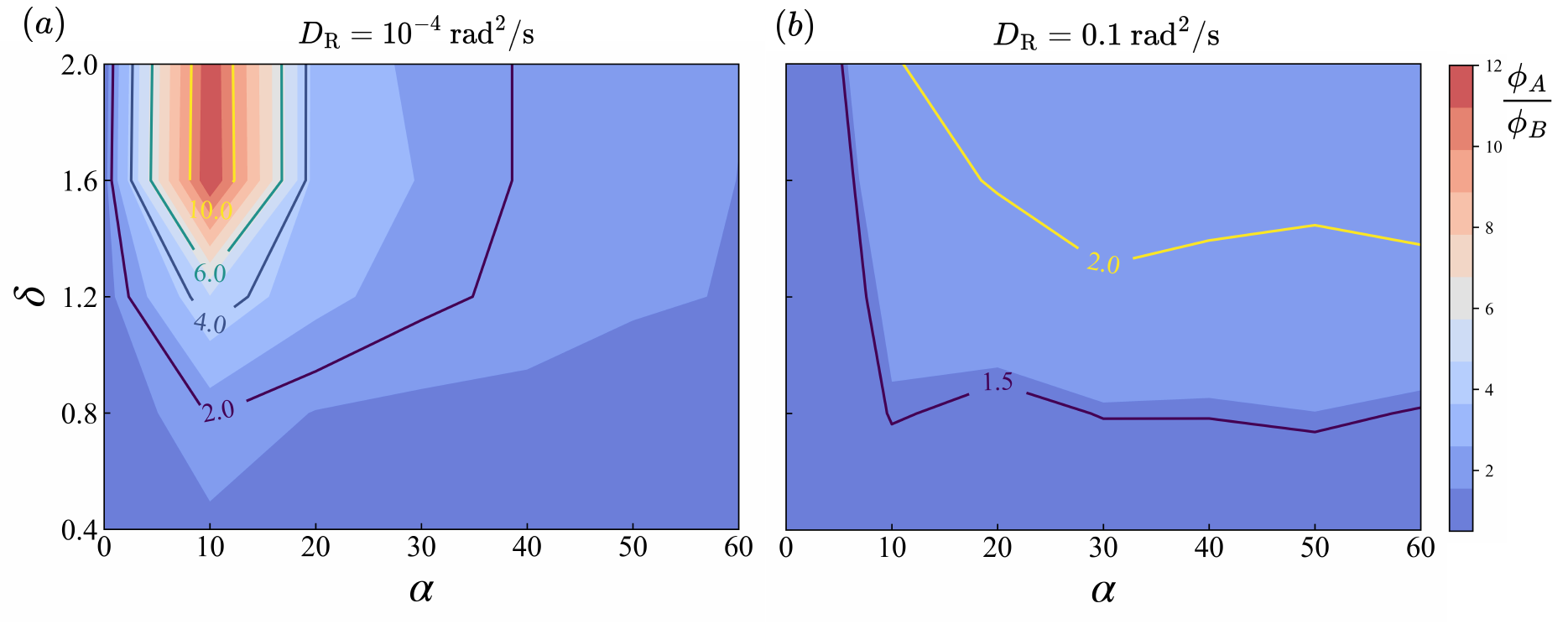}  
        \caption{Separation efficiency $\frac{\phi_A}{\phi_B}$ of a mixture consisting of swimmers with chiralities $\Omega_A$ and $\Omega_B$ with accumulations $\phi_A$ and $\phi_B$ for two values of the rotational diffusion: (a) $D_{\textrm{R}}=10^{-4}$ $\mathrm{rad}^2/\mathrm{s}$ and (b) $D_{\textrm{R}}=0.1$ $\mathrm{rad}^2/\mathrm{s}$. The separation efficiency is displayed as a function of the alignment force, $\alpha$ and the resolution of the separator, $\delta = \Omega_A - \Omega_B$.}
        \label{fig_separator}
\end{figure*}

\subsection{Effect of the obstacle density on accumulation}
Now we proceed to examine the effect of the obstacle density $\rho$. In Fig. \ref{fig_rho} (a,b) $\frac{\phi}{\phi_0}$, the fraction of accumulated swimmers is presented for alignment force strengths $\alpha = 10$ (triangles) and $\alpha = 60$ (circles). 

An increase in the density enhances the effect of the obstacles on the surface accumulation. The accumulation declines with increasing density for $\alpha$ and $\Omega$ where obstacles repel swimmers from the surface ($\frac{\phi}{\phi_0} < 1$ in Fig. \ref{fig_phi} c,d) and increases where orbital trapping dominates the dynamics ($\frac{\phi}{\phi_0} > 1$ in Fig. \ref{fig_phi} c,d). 

As the angular velocity $\Omega$ increases, $\frac{\phi}{\phi_0}$ varies more weakly with the density in Fig. \ref{fig_rho} (a,b). For swimmers with small angular velocities the accumulation declines rapidly for small densities before continuing to slowly decline. Due to the more efficient space exploration of swimmers with a small angular velocity, a lower density of obstacles is sufficient to affect the accumulation. This suggest that obstacles are effective in detaching swimmers with small angular velocities from the surface even at low densities. For swimmers with larger angular velocity, space exploration is less efficient, as increasing the density gradually increases the effect of the obstacles, and larger densities are required to affect the surface accumulation. 

When we compare $D_{\textrm{R}}=10^{-4}$ $\mathrm{rad}^2/\mathrm{s}$  (Fig. \ref{fig_rho} a) and $D_{\textrm{R}}=0.1$ $\mathrm{rad}^2/\mathrm{s}$ (Fig. \ref{fig_rho} b), we note that for $D_{\textrm{R}}=0.1$ $\mathrm{rad}^2/\mathrm{s}$ increasing the obstacle density has a larger effect on $\frac{\phi}{\phi_0}$ for most values of the angular velocity $\Omega$, suggesting that the density dominates the accumulation behavior.

\subsection{Sorting swimmers by angular velocity}
By controlling the accumulation of microswimmers, obstacles may be used to sort swimmers of different angular velocity \cite{sorting_rev, sorting_volpe, sorting_old}. Depending on the angular velocities, the swimmers will tend to accumulate near a surface or remain in the volume with a different probability. For simplicity we will assume that the swimmers in the mixture are identical in all aspects but their angular velocity and study the efficiency of the separation as a function of the alignment force and the different angular velocities of the mixture.

A way to sort swimmers of different angular velocities would be to consider obstacle sizes such that $\Omega<1$ for one fraction (A) of swimmers and $\Omega>1$ for the other (B). We define $\delta = \Omega_A - \Omega_B$ the resolution of the sorter and $\frac{\phi_A}{\phi_B}$ its efficiency. In Fig. \ref{fig_separator} (a,b) the efficiency is displayed for swimmers with $D_{\textrm{R}}=10^{-4}$ $\mathrm{rad}^2/\mathrm{s}$ and $D_{\textrm{R}}=0.1$ $\mathrm{rad}^2/\mathrm{s}$, for different values of the resolution $\delta$ and the alignment force strength $\alpha$. 

We observe that, as the rotational diffusion increases, the ability to sort the swimmers by angular velocity is reduced. For $D_{\textrm{R}}= 10^{-4}$ $\mathrm{rad}^2/\mathrm{s}$ (Fig. \ref{fig_separator} a) the more chiral fraction accumulates up to 11 times as much as the less chiral one, whereas for $D_{\textrm{R}} = 0.1$ $\mathrm{rad}^2/\mathrm{s}$ (Fig. \ref{fig_separator} b) the more chiral fraction accumulates by up to 2.5 times than the less chiral one. 

For $D_{\textrm{R}}=10^{-4}$ $\mathrm{rad}^2/\mathrm{s}$, and $\alpha \approx 10$ the fraction with a higher angular velocity accumulates significantly more than the other fraction and the mixture can be sorted, while for other $\alpha$, the sorter becomes less efficient as $\alpha$ increases. For $D_{\textrm{R}}= 0.1$ $\mathrm{rad}^2/\mathrm{s}$ the alignment force strength needs to be above 20 for the two fractions to accumulate at different rates, and the sorter does not work efficiently.

\section{Conclusion}
We have introduced a 3D model to study the surface accumulation of a microswimmer, induced by steric forces and an effective short-ranged hydrodynamic force that aligns the propagation direction of the swimmer along the nearest boundary (surface or obstacle). Chiral and non-chiral microswimmers were considered, navigating a volume bounded by a bottom and a top surface plane. We introduced obstacles on the surfaces and studied their effect on the surface accumulation.

For smooth surfaces (without obstacles) and structured surfaces (with obstacles) the surface accumulation is enhanced by increasing the alignment force. The alignment force competes with diffusion to prevent the swimmer (chiral and non-chiral) from orienting away from the surface and escaping. When obstacles are added to the surface, the angular velocity of the swimmer is found to strongly affect their accumulation. For a non-chiral swimmer, the obstacles significantly reduce the accumulation when compared to a smooth surface, even when a small fraction of the surface is covered by obstacles. The obstacles prevent a swimmer from aligning with the surface, by guiding it away into the volume, consistent with recent experimental results \cite{forward}. For a chiral swimmer, the ability of obstacles to mitigate the accumulation is reduced with increasing angular velocity. Swimmers with a large angular velocity explore the surface less efficiently, encountering fewer obstacles, which reduces their effect. Moreover, for a chiral swimmer we find that for sufficiently strong alignment forces, the swimmer may be trapped in a trajectory along the boundary of the obstacle \cite{msd5,following2,following3}, resulting in enhanced accumulation. We find that the relevant length scale for trapping is set by the obstacle size, such that as when the swimming orbit is of the size of the obstacle, or smaller, trapping can occur. We further note that, for swimmers that experience a stronger rotational diffusion, the motion becomes more randomized, reducing the effect of the angular velocity on the accumulation.  

The ability of obstacles to impact the accumulation of microswimmers at a surface, may be used to guide the development of materials that selectively hinder or promote the adhesion of microswimmers, e.g., to control the establishment of biofilms in the case of bacteria. Our findings indicate that, by adding obstacles, for non-chiral swimmers, a surface can become more resistant to accumulation, which is in line with studies of bacterial accumulation \cite{surface_rev}. For chiral swimmers, the way obstacles impact surface accumulation of microswimmers is more complex. Our findings suggest that, by selecting the size of the obstacles, we can control the accumulation near the surface. Covering a surface with obstacles could make it more resistant to the accumulation of microswimmers that are only weakly chiral (that swim in orbits larger than the size of the obstacles), but less resistant to strongly chiral swimmers (that swim in orbits with the size of the obstacle, or smaller).

Additionally, the surface accumulation of swimmers of different chiralities, or the tendency of swimmers to accumulate at obstacles that are larger than the radius of their swimming orbit, could be used to design a sorter of microswimmers based on their angular velocity. By selecting from a mixture of microswimmers the most appropriate swimming properties, the efficiency of microswimmers for a specific task e.g. drug-delivery or bioremediation may be improved \cite{sorting_old, sorting_volpe}. Alternatively, chirality-based spermatozoa selection may be employed to select cells with specific swimming traits desirable for artificial fertilization techniques \cite{sperm_separation}. Our results suggest that for chiral swimmers with a rotational diffusion similar to that of bull sperm, effective sorting by angular velocity could be achieved. Future work might include differently shaped obstacles that are known to trap swimmers well to further explore this idea.

In recent studies, for a microswimmer following a convex boundary, the angle between swimmer and boundary, was found to depend on the radius of curvature of the boundary \cite{following2, following3}. By making the alignment force dependent on the radius of curvature we could explicitly include this in our model. Alternatively, in \cite{search} the swimming behavior of an active particle between obstacles, resulting in optimal space exploration, was found to be different for convex and concave obstacles. In the future, the effect of cavities (or concave obstacles), in combination with convex obstacles on surface accumulation of microswimmers could also be explored. In \cite{search} the active particles were confined to a surface, while with the model proposed here would be possible to extend this study to 3D domains.

\section*{Acknowledgements}
We acknowledge financial support by the European Commissions Horizon 2020 research and innovation program under the Marie Sklodowska-Curie Grant Agreement No. 812780 and from the Portuguese Foundation for Science
and Technology (FCT) under Contract no. UIDB/00618/2020 and UIDP/00618/2020. N.A.M.A. and
G.V. also acknowledge support from the UCL MAPS Faculty Visiting Fellowship programme.
%%%END OF MAIN TEXT%%%

%%%REFERENCES%%%
\bibliography{output} %You need to replace "rsc" on this line with the name of your .bib file

\providecommand*{\mcitethebibliography}{\thebibliography}
\csname @ifundefined\endcsname{endmcitethebibliography}
{\let\endmcitethebibliography\endthebibliography}{}
\begin{mcitethebibliography}{48}
\providecommand*{\natexlab}[1]{#1}
\providecommand*{\mciteSetBstSublistMode}[1]{}
\providecommand*{\mciteSetBstMaxWidthForm}[2]{}
\providecommand*{\mciteBstWouldAddEndPuncttrue}
  {\def\EndOfBibitem{\unskip.}}
\providecommand*{\mciteBstWouldAddEndPunctfalse}
  {\let\EndOfBibitem\relax}
\providecommand*{\mciteSetBstMidEndSepPunct}[3]{}
\providecommand*{\mciteSetBstSublistLabelBeginEnd}[3]{}
\providecommand*{\EndOfBibitem}{}
\mciteSetBstSublistMode{f}
\mciteSetBstMaxWidthForm{subitem}
{(\emph{\alph{mcitesubitemcount}})}
\mciteSetBstSublistLabelBeginEnd{\mcitemaxwidthsubitemform\space}
{\relax}{\relax}

\bibitem[Gompper \emph{et~al.}(2020)Gompper, Winkler, Speck, Solon, Nardini, Peruani, Löwen, Golestanian, Kaupp, Alvarez, Kiørboe, Lauga, Poon, DeSimone, Muiños-Landin, Fischer, Söker, Cichos, Kapral, Gaspard, Ripoll, Sagues, Doostmohammadi, Yeomans, Aranson, Bechinger, Stark, Hemelrijk, Nedelec, Sarkar, Aryaksama, Lacroix, Duclos, Yashunsky, Silberzan, Arroyo, and Kale]{roadmap}
G.~Gompper, R.~G. Winkler, T.~Speck, A.~Solon, C.~Nardini, F.~Peruani, H.~Löwen, R.~Golestanian, U.~B. Kaupp, L.~Alvarez, T.~Kiørboe, E.~Lauga, W.~C.~K. Poon, A.~DeSimone, S.~Muiños-Landin, A.~Fischer, N.~A. Söker, F.~Cichos, R.~Kapral, P.~Gaspard, M.~Ripoll, F.~Sagues, A.~Doostmohammadi, J.~M. Yeomans, I.~S. Aranson, C.~Bechinger, H.~Stark, C.~K. Hemelrijk, F.~J. Nedelec, T.~Sarkar, T.~Aryaksama, M.~Lacroix, G.~Duclos, V.~Yashunsky, P.~Silberzan, M.~Arroyo and S.~Kale, \emph{Journal of Physics: Condensed Matter}, 2020, \textbf{32}, 193001\relax
\mciteBstWouldAddEndPuncttrue
\mciteSetBstMidEndSepPunct{\mcitedefaultmidpunct}
{\mcitedefaultendpunct}{\mcitedefaultseppunct}\relax
\EndOfBibitem
\bibitem[Volpe \emph{et~al.}(2022)Volpe, Bechinger, Cichos, Golestanian, Löwen, Sperl, and Volpe]{space}
G.~Volpe, C.~Bechinger, F.~Cichos, R.~Golestanian, H.~Löwen, M.~Sperl and G.~Volpe, \emph{npj Microgravity}, 2022, \textbf{8}, 54\relax
\mciteBstWouldAddEndPuncttrue
\mciteSetBstMidEndSepPunct{\mcitedefaultmidpunct}
{\mcitedefaultendpunct}{\mcitedefaultseppunct}\relax
\EndOfBibitem
\bibitem[Bechinger \emph{et~al.}(2016)Bechinger, Di~Leonardo, Löwen, Reichhardt, Volpe, and Volpe]{rev}
C.~Bechinger, R.~Di~Leonardo, H.~Löwen, C.~Reichhardt, G.~Volpe and G.~Volpe, \emph{Reviews of Modern Physics}, 2016, \textbf{88}, 045006\relax
\mciteBstWouldAddEndPuncttrue
\mciteSetBstMidEndSepPunct{\mcitedefaultmidpunct}
{\mcitedefaultendpunct}{\mcitedefaultseppunct}\relax
\EndOfBibitem
\bibitem[Araújo \emph{et~al.}(2023)Araújo, Janssen, Barois, Boffetta, Cohen, Corbetta, Dauchot, Dijkstra, Durham, Dussutour, Garnier, Gelderblom, Golestanian, Isa, Koenderink, Löwen, Metzler, Polin, Royall, Šarić, Sengupta, Sykes, Trianni, Tuval, Vogel, Yeomans, Zuriguel, Marin, and Volpe]{steering_confinement}
N.~A.~M. Araújo, L.~M.~C. Janssen, T.~Barois, G.~Boffetta, I.~Cohen, A.~Corbetta, O.~Dauchot, M.~Dijkstra, W.~M. Durham, A.~Dussutour, S.~Garnier, H.~Gelderblom, R.~Golestanian, L.~Isa, G.~H. Koenderink, H.~Löwen, R.~Metzler, M.~Polin, C.~P. Royall, A.~Šarić, A.~Sengupta, C.~Sykes, V.~Trianni, I.~Tuval, N.~Vogel, J.~M. Yeomans, I.~Zuriguel, A.~Marin and G.~Volpe, \emph{Soft Matter}, 2023, \textbf{19}, 1695--1704\relax
\mciteBstWouldAddEndPuncttrue
\mciteSetBstMidEndSepPunct{\mcitedefaultmidpunct}
{\mcitedefaultendpunct}{\mcitedefaultseppunct}\relax
\EndOfBibitem
\bibitem[Berke \emph{et~al.}(2008)Berke, Turner, Berg, and Lauga]{hydro}
A.~P. Berke, L.~Turner, H.~C. Berg and E.~Lauga, \emph{Phys. Rev. Lett.}, 2008, \textbf{101}, 038102\relax
\mciteBstWouldAddEndPuncttrue
\mciteSetBstMidEndSepPunct{\mcitedefaultmidpunct}
{\mcitedefaultendpunct}{\mcitedefaultseppunct}\relax
\EndOfBibitem
\bibitem[Li and Tang(2009)]{accumul}
G.~Li and J.~X. Tang, \emph{Phys. Rev. Lett.}, 2009, \textbf{103}, 078101\relax
\mciteBstWouldAddEndPuncttrue
\mciteSetBstMidEndSepPunct{\mcitedefaultmidpunct}
{\mcitedefaultendpunct}{\mcitedefaultseppunct}\relax
\EndOfBibitem
\bibitem[Takagi \emph{et~al.}(2014)Takagi, Palacci, Braunschweig, Shelley, and Zhang]{following1}
D.~Takagi, J.~Palacci, A.~Braunschweig, M.~Shelley and J.~Zhang, \emph{Soft Matter}, 2014, \textbf{10}, 1784--1789\relax
\mciteBstWouldAddEndPuncttrue
\mciteSetBstMidEndSepPunct{\mcitedefaultmidpunct}
{\mcitedefaultendpunct}{\mcitedefaultseppunct}\relax
\EndOfBibitem
\bibitem[Spagnolie \emph{et~al.}(2015)Spagnolie, Moreno-Flores, Bartolo, and Lauga]{following2}
S.~E. Spagnolie, G.~R. Moreno-Flores, D.~Bartolo and E.~Lauga, \emph{Soft Matter}, 2015, \textbf{11}, 3396–3411\relax
\mciteBstWouldAddEndPuncttrue
\mciteSetBstMidEndSepPunct{\mcitedefaultmidpunct}
{\mcitedefaultendpunct}{\mcitedefaultseppunct}\relax
\EndOfBibitem
\bibitem[Sipos \emph{et~al.}(2015)Sipos, Nagy, Di~Leonardo, and Galajda]{following3}
O.~Sipos, K.~Nagy, R.~Di~Leonardo and P.~Galajda, \emph{Phys. Rev. Lett.}, 2015, \textbf{114}, 258104\relax
\mciteBstWouldAddEndPuncttrue
\mciteSetBstMidEndSepPunct{\mcitedefaultmidpunct}
{\mcitedefaultendpunct}{\mcitedefaultseppunct}\relax
\EndOfBibitem
\bibitem[Lauga \emph{et~al.}(2006)Lauga, DiLuzio, Whitesides, and Stone]{whychiral}
E.~Lauga, W.~R. DiLuzio, G.~M. Whitesides and H.~A. Stone, \emph{Biophysical Journal}, 2006, \textbf{90}, 400--412\relax
\mciteBstWouldAddEndPuncttrue
\mciteSetBstMidEndSepPunct{\mcitedefaultmidpunct}
{\mcitedefaultendpunct}{\mcitedefaultseppunct}\relax
\EndOfBibitem
\bibitem[Bukatin \emph{et~al.}(2015)Bukatin, Kukhtevich, Stoop, Dunkel, and Kantsler]{whychiral2}
A.~Bukatin, I.~Kukhtevich, N.~Stoop, J.~Dunkel and V.~Kantsler, \emph{Proceedings of the National Academy of Sciences}, 2015, \textbf{112}, 15904--15909\relax
\mciteBstWouldAddEndPuncttrue
\mciteSetBstMidEndSepPunct{\mcitedefaultmidpunct}
{\mcitedefaultendpunct}{\mcitedefaultseppunct}\relax
\EndOfBibitem
\bibitem[Utada \emph{et~al.}(2014)Utada, Bennett, Fong, Gibiansky, Yildiz, Golestanian, and Wong]{whychiral3}
A.~S. Utada, R.~R. Bennett, J.~C.~N. Fong, M.~L. Gibiansky, F.~H. Yildiz, R.~Golestanian and G.~Wong, \emph{Nature Communications}, 2014, \textbf{5}, 4913\relax
\mciteBstWouldAddEndPuncttrue
\mciteSetBstMidEndSepPunct{\mcitedefaultmidpunct}
{\mcitedefaultendpunct}{\mcitedefaultseppunct}\relax
\EndOfBibitem
\bibitem[Liebchen and Levis(2022)]{rev_liebchen}
B.~Liebchen and D.~Levis, \emph{Chiral Active Matter}, 2022, \url{https://arxiv.org/abs/2207.01923}\relax
\mciteBstWouldAddEndPuncttrue
\mciteSetBstMidEndSepPunct{\mcitedefaultmidpunct}
{\mcitedefaultendpunct}{\mcitedefaultseppunct}\relax
\EndOfBibitem
\bibitem[{Perez Ipi{\~n}a} \emph{et~al.}(2019){Perez Ipi{\~n}a}, {Otte}, {Pontier-Bres}, {Czerucka}, and {Peruani}]{optimal}
E.~{Perez Ipi{\~n}a}, S.~{Otte}, R.~{Pontier-Bres}, D.~{Czerucka} and F.~{Peruani}, \emph{Nature Physics}, 2019, \textbf{15}, 610--615\relax
\mciteBstWouldAddEndPuncttrue
\mciteSetBstMidEndSepPunct{\mcitedefaultmidpunct}
{\mcitedefaultendpunct}{\mcitedefaultseppunct}\relax
\EndOfBibitem
\bibitem[Schultz \emph{et~al.}(2011)Schultz, Bendick, Holm, and Hertel]{biofouling1}
M.~P. Schultz, J.~A. Bendick, E.~R. Holm and W.~M. Hertel, \emph{Biofouling}, 2011, \textbf{27}, 87--98\relax
\mciteBstWouldAddEndPuncttrue
\mciteSetBstMidEndSepPunct{\mcitedefaultmidpunct}
{\mcitedefaultendpunct}{\mcitedefaultseppunct}\relax
\EndOfBibitem
\bibitem[Bixler and Bhushan(2012)]{biofouling2}
G.~D. Bixler and B.~Bhushan, \emph{Philosophical Transactions of the Royal Society A: Mathematical, Physical and Engineering Sciences}, 2012, \textbf{370}, 2381--2417\relax
\mciteBstWouldAddEndPuncttrue
\mciteSetBstMidEndSepPunct{\mcitedefaultmidpunct}
{\mcitedefaultendpunct}{\mcitedefaultseppunct}\relax
\EndOfBibitem
\bibitem[Fiedler \emph{et~al.}(2011)Fiedler, Kolitsch, Kleffner, Henke, Stenger, and Brenner]{copper}
J.~Fiedler, A.~Kolitsch, B.~Kleffner, D.~Henke, S.~Stenger and R.~E. Brenner, \emph{The International Journal of Artificial Organs}, 2011, \textbf{34}, 882--888\relax
\mciteBstWouldAddEndPuncttrue
\mciteSetBstMidEndSepPunct{\mcitedefaultmidpunct}
{\mcitedefaultendpunct}{\mcitedefaultseppunct}\relax
\EndOfBibitem
\bibitem[dong Kang and ming Cao(2012)]{industrial}
G.~dong Kang and Y.~ming Cao, \emph{Water Research}, 2012, \textbf{46}, 584--600\relax
\mciteBstWouldAddEndPuncttrue
\mciteSetBstMidEndSepPunct{\mcitedefaultmidpunct}
{\mcitedefaultendpunct}{\mcitedefaultseppunct}\relax
\EndOfBibitem
\bibitem[Khatoon \emph{et~al.}(2018)Khatoon, McTiernan, Suuronen, Mah, and Alarcon]{medical}
Z.~Khatoon, C.~D. McTiernan, E.~J. Suuronen, T.-F. Mah and E.~I. Alarcon, \emph{Heliyon}, 2018, \textbf{4}, e01067\relax
\mciteBstWouldAddEndPuncttrue
\mciteSetBstMidEndSepPunct{\mcitedefaultmidpunct}
{\mcitedefaultendpunct}{\mcitedefaultseppunct}\relax
\EndOfBibitem
\bibitem[Leonov \emph{et~al.}(2021)Leonov, Flores-Alsina, Gernaey, and Sternberg]{biorefineries}
P.~S. Leonov, X.~Flores-Alsina, K.~V. Gernaey and C.~Sternberg, \emph{Biotechnology Advances}, 2021, \textbf{50}, 107766\relax
\mciteBstWouldAddEndPuncttrue
\mciteSetBstMidEndSepPunct{\mcitedefaultmidpunct}
{\mcitedefaultendpunct}{\mcitedefaultseppunct}\relax
\EndOfBibitem
\bibitem[Gadkari \emph{et~al.}(2022)Gadkari, Bhattacharya, and Shrivastav]{bioremediation}
J.~Gadkari, S.~Bhattacharya and A.~Shrivastav, \emph{Development in Wastewater Treatment Research and Processes}, Elsevier, 2022, pp. 153--173\relax
\mciteBstWouldAddEndPuncttrue
\mciteSetBstMidEndSepPunct{\mcitedefaultmidpunct}
{\mcitedefaultendpunct}{\mcitedefaultseppunct}\relax
\EndOfBibitem
\bibitem[Dade-Robertson \emph{et~al.}(2017)Dade-Robertson, Keren-Paz, Zhang, and Kolodkin-Gal]{template}
M.~Dade-Robertson, A.~Keren-Paz, M.~Zhang and I.~Kolodkin-Gal, \emph{Microbial Biotechnology}, 2017, \textbf{10}, 1157--1163\relax
\mciteBstWouldAddEndPuncttrue
\mciteSetBstMidEndSepPunct{\mcitedefaultmidpunct}
{\mcitedefaultendpunct}{\mcitedefaultseppunct}\relax
\EndOfBibitem
\bibitem[Luan \emph{et~al.}(2018)Luan, Liu, Pihl, {van der Mei}, Liu, Hizal, Choi, Chen, Ren, and Busscher]{surface1}
Y.~Luan, S.~Liu, M.~Pihl, H.~{van der Mei}, J.~Liu, F.~Hizal, C.-H. Choi, H.~Chen, Y.~Ren and H.~Busscher, \emph{Current Opinion in Colloid and Interface Science}, 2018, \textbf{38}, 170--189\relax
\mciteBstWouldAddEndPuncttrue
\mciteSetBstMidEndSepPunct{\mcitedefaultmidpunct}
{\mcitedefaultendpunct}{\mcitedefaultseppunct}\relax
\EndOfBibitem
\bibitem[Shi \emph{et~al.}(2021)Shi, Wang, Cheng, Chen, and Liu]{surface2}
J.~Shi, S.~Wang, X.~Cheng, S.~Chen and G.~Liu, \emph{Journal of Materials Science `I\&' Technology}, 2021, \textbf{70}, 145--155\relax
\mciteBstWouldAddEndPuncttrue
\mciteSetBstMidEndSepPunct{\mcitedefaultmidpunct}
{\mcitedefaultendpunct}{\mcitedefaultseppunct}\relax
\EndOfBibitem
\bibitem[Chinnaraj \emph{et~al.}(2021)Chinnaraj, Jayathilake, Dawson, Ammar, Portoles, Jakubovics, and Chen]{surface3}
S.~B. Chinnaraj, P.~G. Jayathilake, J.~Dawson, Y.~Ammar, J.~Portoles, N.~Jakubovics and J.~Chen, \emph{Journal of Materials Science `I\&' Technology}, 2021, \textbf{81}, 151--161\relax
\mciteBstWouldAddEndPuncttrue
\mciteSetBstMidEndSepPunct{\mcitedefaultmidpunct}
{\mcitedefaultendpunct}{\mcitedefaultseppunct}\relax
\EndOfBibitem
\bibitem[Yang \emph{et~al.}(2022)Yang, Shi, Wang, Chen, Liang, Yang, and Wang]{surface_rev}
K.~Yang, J.~Shi, L.~Wang, Y.~Chen, C.~Liang, L.~Yang and L.-N. Wang, \emph{Journal of Materials Science `I\&' Technology}, 2022, \textbf{99}, 82--100\relax
\mciteBstWouldAddEndPuncttrue
\mciteSetBstMidEndSepPunct{\mcitedefaultmidpunct}
{\mcitedefaultendpunct}{\mcitedefaultseppunct}\relax
\EndOfBibitem
\bibitem[Elgeti and Gompper(2016)]{near_surfaces_rev}
J.~Elgeti and G.~Gompper, \emph{The European Physical Journal Special Topics}, 2016, \textbf{225}, 2333--2352\relax
\mciteBstWouldAddEndPuncttrue
\mciteSetBstMidEndSepPunct{\mcitedefaultmidpunct}
{\mcitedefaultendpunct}{\mcitedefaultseppunct}\relax
\EndOfBibitem
\bibitem[Dehkharghani \emph{et~al.}(2022)Dehkharghani, Waisbord, and Guasto]{experiments1}
A.~Dehkharghani, N.~Waisbord and J.~S. Guasto, \emph{Self-transport of swimming bacteria is impaired by porous microstructure}, 2022\relax
\mciteBstWouldAddEndPuncttrue
\mciteSetBstMidEndSepPunct{\mcitedefaultmidpunct}
{\mcitedefaultendpunct}{\mcitedefaultseppunct}\relax
\EndOfBibitem
\bibitem[Zeitz \emph{et~al.}(2017)Zeitz, Wolff, and Stark]{lorentzgas}
M.~Zeitz, K.~Wolff and H.~Stark, \emph{The European Physical Journal E}, 2017, \textbf{40}, 1--10\relax
\mciteBstWouldAddEndPuncttrue
\mciteSetBstMidEndSepPunct{\mcitedefaultmidpunct}
{\mcitedefaultendpunct}{\mcitedefaultseppunct}\relax
\EndOfBibitem
\bibitem[Chepizhko and Franosch(2019)]{franosh1}
O.~Chepizhko and T.~Franosch, \emph{Soft Matter}, 2019, \textbf{15}, 452--461\relax
\mciteBstWouldAddEndPuncttrue
\mciteSetBstMidEndSepPunct{\mcitedefaultmidpunct}
{\mcitedefaultendpunct}{\mcitedefaultseppunct}\relax
\EndOfBibitem
\bibitem[Chepizhko and Franosch(2020)]{franosh2}
O.~Chepizhko and T.~Franosch, \emph{New Journal of Physics}, 2020, \textbf{22}, 073022\relax
\mciteBstWouldAddEndPuncttrue
\mciteSetBstMidEndSepPunct{\mcitedefaultmidpunct}
{\mcitedefaultendpunct}{\mcitedefaultseppunct}\relax
\EndOfBibitem
\bibitem[van Roon \emph{et~al.}(2022)van Roon, Volpe, Telo~da Gama, and Araújo]{chiral}
D.~M. van Roon, G.~Volpe, M.~M. Telo~da Gama and N.~A.~M. Araújo, \emph{Soft Matter}, 2022, \textbf{18}, 6899--6906\relax
\mciteBstWouldAddEndPuncttrue
\mciteSetBstMidEndSepPunct{\mcitedefaultmidpunct}
{\mcitedefaultendpunct}{\mcitedefaultseppunct}\relax
\EndOfBibitem
\bibitem[Makarchuk \emph{et~al.}(2019)Makarchuk, Braz, Araujo, Ciric, and Volpe]{forward}
S.~Makarchuk, V.~Braz, N.~Araujo, L.~Ciric and G.~Volpe, \emph{Nature Communications}, 2019, \textbf{10}, 4110\relax
\mciteBstWouldAddEndPuncttrue
\mciteSetBstMidEndSepPunct{\mcitedefaultmidpunct}
{\mcitedefaultendpunct}{\mcitedefaultseppunct}\relax
\EndOfBibitem
\bibitem[Brown \emph{et~al.}(2016)Brown, Vladescu, Dawson, Vissers, Schwarz-Linek, Lintuvuori, and Poon]{lattice}
A.~T. Brown, I.~D. Vladescu, A.~Dawson, T.~Vissers, J.~Schwarz-Linek, J.~S. Lintuvuori and W.~C.~K. Poon, \emph{Soft Matter}, 2016, \textbf{12}, 131--140\relax
\mciteBstWouldAddEndPuncttrue
\mciteSetBstMidEndSepPunct{\mcitedefaultmidpunct}
{\mcitedefaultendpunct}{\mcitedefaultseppunct}\relax
\EndOfBibitem
\bibitem[Weber \emph{et~al.}(2019)Weber, Bahrs, Alirezaeizanjani, Zhang, Beta, and Zaburdaev]{lattice2}
A.~Weber, M.~Bahrs, Z.~Alirezaeizanjani, X.~Zhang, C.~Beta and V.~Zaburdaev, \emph{Frontiers in Physics}, 2019, \textbf{7}, 148\relax
\mciteBstWouldAddEndPuncttrue
\mciteSetBstMidEndSepPunct{\mcitedefaultmidpunct}
{\mcitedefaultendpunct}{\mcitedefaultseppunct}\relax
\EndOfBibitem
\bibitem[Brun-Cosme-Bruny \emph{et~al.}(2019)Brun-Cosme-Bruny, Bertin, Coasne, Peyla, and Rafaï]{lattice3}
M.~Brun-Cosme-Bruny, E.~Bertin, B.~Coasne, P.~Peyla and S.~Rafaï, \emph{The Journal of Chemical Physics}, 2019, \textbf{150}, 104901\relax
\mciteBstWouldAddEndPuncttrue
\mciteSetBstMidEndSepPunct{\mcitedefaultmidpunct}
{\mcitedefaultendpunct}{\mcitedefaultseppunct}\relax
\EndOfBibitem
\bibitem[Kurzthaler \emph{et~al.}(2021)Kurzthaler, Mandal, Bhattacharjee, L{\"o}wen, Dattaf, and Stone]{3d}
C.~Kurzthaler, S.~Mandal, T.~Bhattacharjee, H.~L{\"o}wen, S.~Dattaf and H.~Stone, \emph{Nature Communications}, 2021, \textbf{12}, 7088\relax
\mciteBstWouldAddEndPuncttrue
\mciteSetBstMidEndSepPunct{\mcitedefaultmidpunct}
{\mcitedefaultendpunct}{\mcitedefaultseppunct}\relax
\EndOfBibitem
\bibitem[Bianchi \emph{et~al.}(2017)Bianchi, Saglimbeni, and Di~Leonardo]{leonardo1}
S.~Bianchi, F.~Saglimbeni and R.~Di~Leonardo, \emph{Phys. Rev. X}, 2017, \textbf{7}, 011010\relax
\mciteBstWouldAddEndPuncttrue
\mciteSetBstMidEndSepPunct{\mcitedefaultmidpunct}
{\mcitedefaultendpunct}{\mcitedefaultseppunct}\relax
\EndOfBibitem
\bibitem[Bianchi \emph{et~al.}(2019)Bianchi, Saglimbeni, Frangipane, Dell{'}Arciprete, and Di~Leonardo]{leonardo2}
S.~Bianchi, F.~Saglimbeni, G.~Frangipane, D.~Dell{'}Arciprete and R.~Di~Leonardo, \emph{Soft Matter}, 2019, \textbf{15}, 3397--3406\relax
\mciteBstWouldAddEndPuncttrue
\mciteSetBstMidEndSepPunct{\mcitedefaultmidpunct}
{\mcitedefaultendpunct}{\mcitedefaultseppunct}\relax
\EndOfBibitem
\bibitem[Drescher \emph{et~al.}(2011)Drescher, Dunkel, Cisneros, Ganguly, and Goldstein]{shortranged}
K.~Drescher, J.~Dunkel, L.~H. Cisneros, S.~Ganguly and R.~E. Goldstein, \emph{Proceedings of the National Academy of Sciences}, 2011, \textbf{108}, 10940 -- 10945\relax
\mciteBstWouldAddEndPuncttrue
\mciteSetBstMidEndSepPunct{\mcitedefaultmidpunct}
{\mcitedefaultendpunct}{\mcitedefaultseppunct}\relax
\EndOfBibitem
\bibitem[Li and Tang(2009)]{bullsperm}
G.~Li and J.~X. Tang, \emph{Phys. Rev. Lett.}, 2009, \textbf{103}, 078101\relax
\mciteBstWouldAddEndPuncttrue
\mciteSetBstMidEndSepPunct{\mcitedefaultmidpunct}
{\mcitedefaultendpunct}{\mcitedefaultseppunct}\relax
\EndOfBibitem
\bibitem[Whitesides \emph{et~al.}(2001)Whitesides, Ostuni, Takayama, Jiang, and Ingber]{obstacle_size}
G.~M. Whitesides, E.~Ostuni, S.~Takayama, X.~Jiang and D.~E. Ingber, \emph{Annual Review of Biomedical Engineering}, 2001, \textbf{3}, 335--373\relax
\mciteBstWouldAddEndPuncttrue
\mciteSetBstMidEndSepPunct{\mcitedefaultmidpunct}
{\mcitedefaultendpunct}{\mcitedefaultseppunct}\relax
\EndOfBibitem
\bibitem[Laird(2011)]{sorting_rev}
T.~Laird, \emph{Organic Process Research \& Development}, 2011, \textbf{15}, 946--946\relax
\mciteBstWouldAddEndPuncttrue
\mciteSetBstMidEndSepPunct{\mcitedefaultmidpunct}
{\mcitedefaultendpunct}{\mcitedefaultseppunct}\relax
\EndOfBibitem
\bibitem[Mijalkov and Volpe(2013)]{sorting_volpe}
M.~Mijalkov and G.~Volpe, \emph{Soft Matter}, 2013, \textbf{9}, 6376--6381\relax
\mciteBstWouldAddEndPuncttrue
\mciteSetBstMidEndSepPunct{\mcitedefaultmidpunct}
{\mcitedefaultendpunct}{\mcitedefaultseppunct}\relax
\EndOfBibitem
\bibitem[Marcos \emph{et~al.}(2009)Marcos, Fu, Powers, and Stocker]{sorting_old}
Marcos, H.~C. Fu, T.~R. Powers and R.~Stocker, \emph{Physical Review Letters}, 2009, \textbf{102}, year\relax
\mciteBstWouldAddEndPuncttrue
\mciteSetBstMidEndSepPunct{\mcitedefaultmidpunct}
{\mcitedefaultendpunct}{\mcitedefaultseppunct}\relax
\EndOfBibitem
\bibitem[van Teeffelen and L\"owen(2008)]{msd5}
S.~van Teeffelen and H.~L\"owen, \emph{Physical Review Letters E}, 2008, \textbf{78}, 020101\relax
\mciteBstWouldAddEndPuncttrue
\mciteSetBstMidEndSepPunct{\mcitedefaultmidpunct}
{\mcitedefaultendpunct}{\mcitedefaultseppunct}\relax
\EndOfBibitem
\bibitem[Oseguera-López \emph{et~al.}(2019)Oseguera-López, Ruiz-Díaz, Ramos-Ibeas, and Pérez-Cerezales]{sperm_separation}
I.~Oseguera-López, S.~Ruiz-Díaz, P.~Ramos-Ibeas and S.~Pérez-Cerezales, \emph{Frontiers in Cell and Developmental Biology}, 2019, \textbf{7}, year\relax
\mciteBstWouldAddEndPuncttrue
\mciteSetBstMidEndSepPunct{\mcitedefaultmidpunct}
{\mcitedefaultendpunct}{\mcitedefaultseppunct}\relax
\EndOfBibitem
\bibitem[Volpe and Volpe(2017)]{search}
G.~Volpe and G.~Volpe, \emph{Proceedings of the National Academy of Sciences}, 2017, \textbf{114}, 11350--11355\relax
\mciteBstWouldAddEndPuncttrue
\mciteSetBstMidEndSepPunct{\mcitedefaultmidpunct}
{\mcitedefaultendpunct}{\mcitedefaultseppunct}\relax
\EndOfBibitem
\end{mcitethebibliography}
\bibliographystyle{output} %the RSC's .bst file

\end{document}